\documentclass[aps,prb,floats,floatsfix,twocolumn,showpacs]{revtex4-1}

\usepackage{dcolumn} 
\usepackage{bm} 
\usepackage{amsthm}
\usepackage{latexsym}
\usepackage{float}
\usepackage{amsfonts}
\usepackage{amsmath}
\usepackage{amssymb}
\usepackage{color,graphicx}

\begin{document}
\title{Homogeneous Linewidth Narrowing of the Charged Exciton via Nuclear Spin Screening in an InAs/GaAs Quantum Dot Ensemble}

\author{G.~Moody}
\affiliation{National Institute of Standards and Technology, Boulder CO 80305}
\author{M.~Feng}
\affiliation{National Institute of Standards and Technology, Boulder CO 80305}
\author{C.~McDonald}
\affiliation{National Institute of Standards and Technology, Boulder CO 80305}
\author{R.~P.~Mirin}
\affiliation{National Institute of Standards and Technology, Boulder CO 80305}
\author{K.~L.~Silverman}
\email{kevin.silverman@nist.gov}
\affiliation{National Institute of Standards and Technology, Boulder CO 80305}

\begin{abstract}
In semiconductor quantum dots, the electron hyperfine interaction with the nuclear spin bath is the leading source of spin decoherence at cryogenic temperature.  Using high-resolution two-color differential transmission spectroscopy, we demonstrate that such electron-nuclear coupling also imposes a lower limit for the positively charged exciton dephasing rate, $\gamma$, in an ensemble of InAs/GaAs quantum dots.  We find that $\gamma$ is sensitive to the strength of the hyperfine interaction, which can be controlled through the application of an external magnetic field in the Faraday configuration.  At zero applied field, strong electron-nuclear coupling induces additional dephasing beyond the radiative limit and $\gamma = 230$ MHz ($0.95$ $\mu$eV).  Screening of the hyperfine interaction is achieved for an external field of $\approx 1$ T, resulting in $\gamma = 172$ MHz ($0.71$ $\mu$eV) limited only by spontaneous recombination.  On the other hand, application of a Voigt magnetic field mixes the spin eigenstates, which increases $\gamma$ by up to $75 \%$.  These results are reproduced with a simple and intuitive model that captures the essential features of the electron hyperfine interaction and its influence on $\gamma$.
\end{abstract}

\date{\today}
\maketitle

\section{Introduction}
\label{Introduction}

The spin of a charge carrier confined in a semiconductor quantum dot (QD) is an excellent platform for realizing a solid-state qubit, a fundamental building block for quantum information devices\cite{Ramsay2010}.  Fast initialization, manipulation, and readout of the spin qubit can be achieved through optical excitation of an intermediate charged exciton state\cite{Bracker2005,Xu2007,Gerardot2008}, comprised of the confined carrier Coulomb-bound to an optically excited electron-hole pair.  Such a hybrid scheme combines the long coherence time of carrier spins\cite{Kroutvar2004,Heiss2007} with the picosecond gate time of excitonic qubits \cite{Press2008,Godden2012}, which can be leveraged for robust and scalable qubit implementation.  A critical parameter for realizing spin qubits in QDs is the charged exciton optical coherence time, which is inversely proportional to the dephasing rate, $\gamma$.  Since $\gamma$ ultimately limits the timescale during which operations based on coherent light-matter interactions can be performed, understanding how it is affected by coupling of the charged exciton to its environment -- and how to control such coupling -- is imperative.

Despite its importance, $\gamma$ has been challenging to measure for semiconductor QDs.  Photoluminescence spectroscopy has been the primary tool for characterizing the transition energy and lineshape, which requires isolation of single QDs to avoid inhomogeneous broadening\cite{Gammon1996,Dalgarno2008}.  However, single dot experiments, which require long signal integration times, still suffer from spectral diffusion effects that mask the intrinsic recombination dynamics\cite{Hogele2004,Berthelot2006,Gerardot2008}.  These limitations can be avoided by investigating QD ensembles using nonlinear optical spectroscopy techniques, such as spectral hole burning \cite{Bonadeo2000,Palinginis2003,Berry2006}, four-wave mixing \cite{Borri2007}, or coherent multi-dimensional spectroscopy \cite{Moody2013a,Moody2013b}.  An ultralong coherence time up to $\sim 1.5$ ns, corresponding to a sub-$\mu$eV dephasing rate, has been measured for the positively-charged exciton in InAs QDs at cryogenic temperature\cite{Cesari2010}.  Such a long coherence time indicates almost complete absence of pure dephasing effects arising from exciton-exciton and exciton-phonon interactions.  In this limit, $\gamma$ is governed by the population relaxation dynamics, which are often assumed to be purely radiative.

The three-dimensional confinement of charge carriers that suppresses pure dephasing effects, compared to bulk materials, unfortunately enhances the hyperfine interaction of the carrier spins with the $\approx 10^5$ nuclear spins that preside in the QD \cite{Merkulov2002,SpinPhysics2008}.  In this work, we demonstrate that signatures of carrier-nuclear spin interactions are imprinted on the charged exciton coherence dynamics, consequently affecting $\gamma$.  For electron spins, the hyperfine interaction, stemming from Fermi contact coupling, has been identified as the dominant electron spin dephasing mechanism at low temperature in the absence of external fields \cite{Khaetskii2002}.  For hole spins, the $p$-like symmetry of the valence band states renders the hole-nuclear interaction relatively inefficient, resulting in a smaller hole-spin relaxation rate compared to the electron\cite{Heiss2007,Eble2009,Fallahi2010}.  This presents an advantage for using positively charged excitons for spin qubits: hole-spin state initialization can be achieved by optical spin pumping through the intermediate positively charged exciton state with near unity fidelity at zero external magnetic field \cite{Gerardot2008}.  Previous studies have primarily focused on examining how the hyperfine interaction leads to carrier spin decoherence and limits spin state initialization fidelity.  Hyperfine-mediated spin dephasing effects on $\gamma$, on the other hand, have so far been unexplored.

Motivated by this, we use resonant two-color differential transmission spectroscopy to measure $\gamma$ of the positively charged exciton in a small ensemble of InAs/GaAs QDs in both the Faraday and Voigt magnetic field configurations.  This technique enables high-resolution measurements of $\gamma$ in the absence of exciton- and phonon-mediated interactions, which isolates contributions from the electron hyperfine interaction.  At zero external magnetic field, we measure a sub-$\mu$eV dephasing rate, which \textit{decreases} by $\approx 25 \%$ for a Faraday field of $\approx 1$ T.  This result is reproduced using a simple and intuitive model\cite{Kroner2008} in which the strength of the electron hyperfine interaction is determined by the energetic overlap of the electron spin states.  As the spin states are Zeeman shifted with increasing magnetic field, the electron spin flip probability decreases, resulting in a weaker hyperfine contribution to $\gamma$.  For fields $\gtrsim 1$ T, we obtained a radiatively-limited dephasing rate $\gamma = 172$ MHz ($h\gamma = 0.71$ $\mu$eV).  On the other hand, application of a Voigt field mixes the hole and charged exciton spin states, enabling the ``forbidden" dark state transitions.  As the field strength is increased up to 3 T, $\gamma$ increases by $\approx 75 \%$, corresponding to a dipole moment ratio between the dark and bright states of $\approx 0.85$.  These results demonstrate that even in the absence of carrier- and phonon-mediated interactions, $\gamma$ is not limited by spontaneous recombination.  Instead, the strong hyperfine interaction in semiconductor QDs enhances $\gamma$ beyond the radiative limit, which can be suppressed by the application of a moderate magnetic field.

\section{Experimental Methods and Sample}
\subsection{Sample Properties and Characterization}
\label{Sample}

\begin{figure}[h]
\centering
\includegraphics[width=0.99\columnwidth]{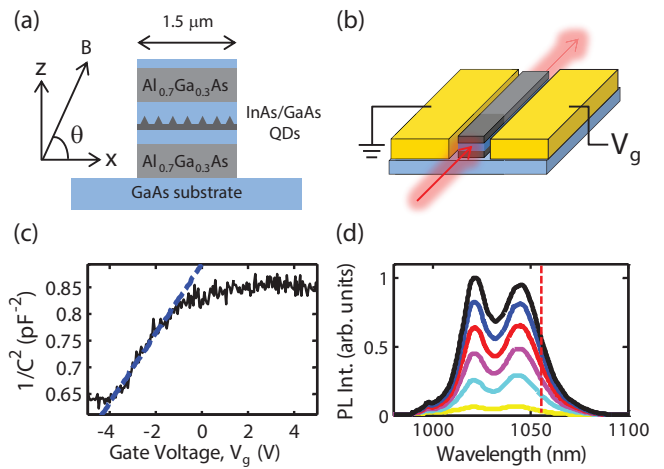}
\caption{The sample studied is a single layer of InAs/GaAs self-assembled QDs in a 1.5 $\mu$m wide ridge waveguide structure, shown in the schematic diagrams in (a) and (b).  Gold electrodes separated by 2 $\mu$m are patterned on both sides of the waveguide for capacitance-voltage (C-V) measurements of the device.  A magnetic field up to 3 T can be applied along either the growth direction (Faraday configuration, $\theta = \pi/2$) or in the plane of the QDs (Voigt configuration, $\theta = 0$), as indicated in (a).  Room temperature C-V measurements reveal that the sample is unintentionally $p$-doped, determined by the slope of the dashed line fit to the $1/C^2$ vs. gate voltage ($V_g$) data shown in (c).  Panel (d) shows photoluminescence spectra of positively-charged excitons taken at 4.2 K using 980 nm excitation with increasing power from 50 $\mu$W to 3.75 mW.  The vertical dashed line indicates the laser wavelength for the differential transmission experiments.}
\label{fig1}
\end{figure}

The investigated sample consists of a single layer of nominally undoped In(Ga)As/GaAs self-assembled QDs embedded in a 1.5 $\mu$m-wide and 1 mm-long ridge waveguide structure, shown in the schematic diagrams in Figs. \ref{fig1}(a) and \ref{fig1}(b).  The waveguide ridge surfaces are passivated with 50 nm of SiN$_x$, after which $500$ nm thick Au electrodes are patterned on either side of the waveguide.  The device is electrically insulated by a thick benzocyclobutene coating.  The waveguide serves to confine the laser fields to the QD core along the $x$-direction through the semiconductor/SiN$_x$ refractive index mismatch and along the $z$-direction by the 1.1 $\mu$m-thick Al$_{0.7}$Ga$_{0.3}$As cladding layers, effectively increasing the light-matter interaction length of laser fields propagating along the $y$-direction.

Although the sample is nominally undoped, residual impurities present during the epitaxial growth process can introduce free carriers into the wetting layer and QDs.  The device geometry \cite{Vasileska1997,SchredToolkit,EN1} enables straightforward electrical characterization of the doping properties of the semiconductor material through room temperature capacitance-voltage (C-V) measurements.  C-V data is obtained by applying a 5 mV, 20 kHz AC voltage superimposed onto a DC bias to the gate electrode ($V_g$).  The DC voltage is swept from +5V to -5V at a 0.05 V/s sweep rate to allow sampling of the doping profile at different depths in the device.  Plotting $1/C^2$ vs. $V_g$, shown by the solid line in Fig. \ref{fig1}(c), yields information regarding the doping characteristics since the material doping density, $N_p$, is inversely proportional to the slope of curve in depletion.  $N_p$ is related to the material properties and device geometry through \cite{Schroder2006}
\begin{equation}
N_p = \left|\frac{-2}{q \cdot \epsilon_s \cdot A_c^2 \cdot \frac{d\left(1/C^2\right)}{dV_g}}\right|,
\label{eqn_dopingdensity}
\end{equation}
where $q \approx 1.6 \times 10^{-19}$ C is the electric charge, $\epsilon_s \approx 11 \cdot 8.85 \times 10^{-14}$ F/cm is the semiconductor dielectric constant \cite{EN2}, and $A_c = 500$ nm $\times$ 1 mm $=5 \times 10^{-6}$ cm$^2$ is the capacitor area.  Using $d\left(1/C^2\right)/dV_g$ given by the dashed line in Fig. \ref{fig1}(c), we find that $N_p \approx 1\times10^{17}$ cm$^{-3}$.  The large $N_p$ likely arises from impurities introduced by the waveguide cladding layers with large Al mole-fraction.  Since the sample was grown to yield a high density of QDs ($10^{10} - 10^{11}$ cm$^{-2}$), we estimate that on average each dot contains a single hole.  This point is further supported by the sub-$\mu$eV linewidth presented in section \ref{Results}, which precludes the presence of additional holes in the QDs since multiply-charged excitons exhibit a dephasing rate on the order of $100$ $\mu$eV or larger \cite{Cesari2010}.

Low temperature photoluminescence spectra are shown in Fig. \ref{fig1}(d) for excitation at 980 nm with increasing power from 50 $\mu$W to 3.75 mW.  The spectra feature two peaks centered at 1045 nm and 1020 nm.  We attribute the lower energy peak to the inhomogeneously broadened ground state of the positively charged exciton, $X^+$, and the higher energy peak corresponds to the first excited state.  For the differential transmission experiments, the excitation lasers are tuned to 1055 nm to ensure that only the ground state transitions of $X^+$ are probed.  The $X^+$ peak assignment is also confirmed by the absence of fine-structure splitting in single dot photoluminescence spectra from other samples epitaxially grown using the same equipment under similar conditions (data not shown).

\subsection{High-Resolution Differential Transmission Spectroscopy}
\label{SHB}

For the optical spectroscopy experiments, the sample is mounted in a transmission confocal microscope setup that is hermetically sealed in a stainless steel insert with 15 Torr of ultra-high purity helium gas.  The sample insert is placed into a liquid helium bath cryostat, where the sample resides at the center point of split coil superconducting magnet, enabling the application of an external magnetic field up to 3 T at a sample temperature of 4.2 K.  Both Faraday ($\theta = \pi/2$ in Fig. \ref{fig1}(a)) and Voigt ($\theta = 0$) magnetic field configurations are possible by rotating the sample insert about the axis of the cryostat.  Light is coupled into and out of the cryostat using single-mode polarization-maintaining fiber.  Coupling to the waveguide is achieved using 0.55 numerical aperture objective lenses each mounted on a three-axis piezoelectric nanopositioner.

\begin{figure}[h]
\centering
\includegraphics[width=0.9\columnwidth]{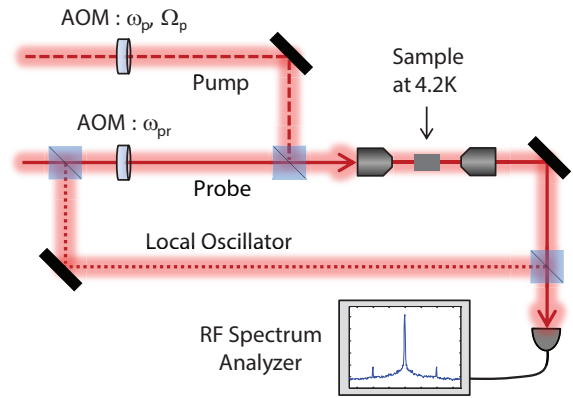}
\caption{Schematic diagram of the high-resolution differential transmission experimental setup.  Acousto-optic modulators (AOMs) shift the narrow-linewidth pump and probe laser beam frequencies by $\omega_p$ and $\omega_{pr}$, respectively, and the pump amplitude is modulated at a frequency $\Omega_p$.  The pump and probe beams  transmitted through the waveguide are heterodyned with a local oscillator.  The differential change in the probe transmission is recorded by a radio frequency spectrum analyzer at the beatnote frequencies $\omega_{pr}\pm\Omega_p$ vs. the pump-probe detuning frequency.}
\label{fig2}
\end{figure}

A schematic diagram of the differential transmission experimental setup is shown in Fig. \ref{fig2}.  Two narrow-frequency laser beams are coupled into the waveguide through one of the optical fibers.  Each laser has a linewidth $< 200$ kHz, and the heterodyne linewidth between the lasers is $\approx$ 4 MHz for a 500 ms integration time.  Using such narrow frequency lasers enables us to probe an estimated $\approx$ 10 QDs, which eliminates effects arising from QD inhomogeneities while still allowing for acceptable signal-to-noise.  One of the lasers remains at a fixed frequency and acts as a pump that burns a spectral hole in the ground state inhomogeneous distribution.  The second laser serves as a weak probe beam that is swept in frequency relative to the pump to map out the homogeneous lineshape of $X^+$.  Discrimination of the probe from the pump is achieved through an optical heterodyne technique in which the probe laser is split into two beams.  The frequency of the probe beam is shifted by $\omega_{pr} = 61$ MHz with an acousto-optic modulator (AOM) and is transmitted through the waveguide.  The other beam is routed around the sample and acts as a local oscillator reference for the probe.  A fast photodiode and radio-frequency spectrum analyzer are used to record the heterodyne beatnote between the probe and local oscillator at $\omega_{pr} = 61$ MHz, which is free from any DC contamination from the pump beam.

The amplitude of the pump laser is modulated at a frequency $\Omega_p = 5$ kHz using another AOM, which subsequently modulates the excited state population density in the QDs.  The resulting probe differential transmission signal appears as sidebands on the probe/local oscillator beat note at $\omega_{pr} \pm \Omega_p$.  Using a radio-frequency spectrum analyzer, the sideband amplitudes are averaged and recorded as a function of the pump-probe detuning frequency.  Heterodyning of the probe with the local oscillator also provides the necessary amplification to avoid power broadening of the homogeneous lineshape, discussed in section \ref{PowerBroadening}.  In all measurements, the pump and probe beams are collinearly polarized along the $x$-direction.  Consistent with previous experiments of excitons in similar waveguide structures \cite{Silverman2003,Berry2006}, no differential transmission signal is observed for either the pump or probe polarized along the growth direction $z$.  The probe power is kept at 2 pW to avoid self-broadening effects, while the pump power is either varied from 2 pW to 2 nW for the power broadening measurements or fixed at 10 pW for the magnetic field measurements.

\subsection{Energy Level Scheme for the $X^+$ Differential Transmission Experiments}
\label{EnergyScheme}

\begin{figure}[h]
\centering
\includegraphics[width=0.6\columnwidth]{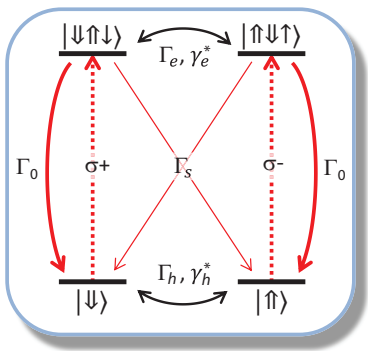}
\caption{Energy level diagram of the ground state charged exciton transitions.  The system ground state consists of a spin down ($\left|\Downarrow\right>$) and spin up ($\left|\Uparrow\right>$) heavy-hole that are coupled by a spin flip relaxation rate $\Gamma_h$ and pure spin dephasing rate $\gamma^*_h$.  The positively charged exciton state $\left|\Uparrow\Downarrow\uparrow\right>$ ($\left|\Downarrow\Uparrow\downarrow\right>$), consisting of a spin-up (-down) electron $\left|\uparrow\right>$ ($\left|\downarrow\right>$) Coulomb bound to the holes in a spin singlet state, is accessible via $\sigma-$ ($\sigma+$) polarized excitation from the $\left|\Uparrow\right>$ ($\left|\Downarrow\right>$) hole state.  The charged exciton states can relax to the same hole ground state through dipole-allowed spontaneous recombination at a rate $\Gamma_0$.  Electron spin relaxation at a rate $\Gamma_e$ and pure dephasing at a rate $\gamma^*_e$ couple the excited states as well as enable the diagonal ``dipole-forbidden" recombination pathways through a second-order process characterized by a rate $\Gamma_s$.}
\label{fig3}
\end{figure}

A description is given here to introduce the relevant $X^+$ transitions and their dephasing rates.  The charged exciton can be represented by a four-level system, shown in Fig. \ref{fig3} for zero external magnetic field.  For the following description, the sample growth $z$-direction is chosen as the direction onto which the spins are projected.  The ground state consists of either a spin-up ($\left|\Uparrow\right> = +3/2$) or a spin-down ($\left|\Downarrow\right> = -3/2$) heavy-hole. Optical selection rules allow for two bright circularly polarized transitions that couple the hole and charged exciton spin states: $\left|\Uparrow\right> \leftrightarrow \left|\Uparrow\Downarrow\uparrow\right>$ for $\sigma-$ polarization and $\left|\Downarrow\right> \leftrightarrow \left|\Downarrow\Uparrow\downarrow\right>$ for $\sigma+$ polarization.  Each $X^+$ state consists of two holes in a singlet spin state and an unpaired electron (indicated by the thin arrows) in either the spin-up ($\left|\Uparrow\Downarrow\uparrow\right> = +1/2$) or spin-down ($\left|\Downarrow\Uparrow\downarrow\right> = -1/2$) state.  The dephasing rate of either optically allowed transition is given by $\gamma = \Gamma/2 + \gamma^*$, which is inversely proportional to the $X^+$ coherence time, $T_2$.  The population relaxation rate, $\Gamma$, describes inelastic processes associated with spontaneous recombination and population transfer and is inversely proportional to the longitudinal relaxation time, $T_1$.  Pure dephasing processes that affect only the relative phases of the eigenstates of the transition without resulting in energy transfer are given by the rate $\gamma^*$.

Hyperfine contributions to $\gamma$ are revealed in the low-temperature differential transmission experiments since additional sources of linewidth broadening arising from carrier-carrier and carrier-phonon interactions, spectral diffusion from trapped charges, and size-dependent inhomogeneities are suppressed.  As the electron simultaneously interacts with $\approx 10^5$ nuclear spins in the QD, one can approximate the electron-nuclear coupling using a mean-field approach for which the average nuclear spin polarization acts like an average effective macroscopic Overhauser magnetic field, \textbf{\textit{B}}$_N$, with fluctuations given by $\delta\textbf{\textit{B}}_N$. In the absence of dynamic nuclear spin polarization, \textbf{\textit{B}}$_N$ $= 0$ and $\delta\textbf{\textit{B}}_N = \textbf{\textit{B}}^{max}_N/\sqrt{N} = 20 - 40$ mT, where $N$ is the number of nuclear spins in the QD and $\textbf{\textit{B}}^{max}_N$ is the maximum Overhauser field corresponding to $100\%$ nuclear spin polarization\cite{Urb2013}.  The nuclear spin polarization can be considered ``frozen" since the spin correlation time is significantly longer than the relevant timescales for electron spin and $X^+$ dephasing \cite{Merkulov2002}.

The Fermi contact hyperfine interaction of the electron with the nuclear spin bath destroys the $X^+$ optical coherence by providing auxiliary recombination pathways in addition to spontaneous recombination.  Upon optical excitation of either dipole-allowed $X^+$ transition, the electron spin coherently precesses about the ``frozen" $\delta\textbf{\textit{B}}_N$ with a coherence dephasing rate given by $\gamma_{e} = \Gamma_{e}/2 + \gamma^*_{e}$, where $\Gamma_{e}$ is the longitudinal spin relaxation rate and $\gamma^*_{e}$ is the spin coherence pure dephasing rate.  Electron spin precession coherently couples the $\left|\Uparrow\Downarrow\uparrow\right>$ and $\left|\Downarrow\Uparrow\downarrow\right>$ states, which can contribute to the $X^+$ dephasing rate $\gamma$ either through $\Gamma$ or $\gamma^*$.  Additionally, electron-nuclear spin coupling relaxes the optical selection rules so that the diagonal transitions $\left|\Downarrow\right> \leftrightarrow \left|\Uparrow\Downarrow\uparrow\right>$ and $\left|\Uparrow\right> \leftrightarrow \left|\Downarrow\Uparrow\downarrow\right>$ are no longer ``forbidden".  Recombination along these pathways is a second order process that requires simultaneously an electron spin flip event and emission of a photon with a characteristic rate $\Gamma_s$.  The diagonal transitions are also weakly allowed through heavy-hole--light-hole valence band mixing \cite{Dreiser2008,Kroner2008}, although the rate of this process is only $\approx 1\%$ of $\Gamma_0$.  In principle, the hole hyperfine interaction will also couple the hole spin states with a spin flip relaxation rate $\Gamma_h$ and pure spin dephasing rate  $\gamma^*_h$.  For holes, however, the contact hyperfine interaction is suppressed due to the $p$-like symmetry of the hole wavefunction.  Dipole-dipole coupling is the leading contribution to hole-nuclear coupling, but it is an order of magnitude weaker compared to the electron hyperfine interaction \cite{Heiss2007,Eble2009,Chekhovich2011}.

\section{Results and Discussion}
\label{Results}

\subsection{Power Broadening of the Homogeneous Lineshape}
\label{PowerBroadening}

\begin{figure}[h]
\centering
\includegraphics[width=1\columnwidth]{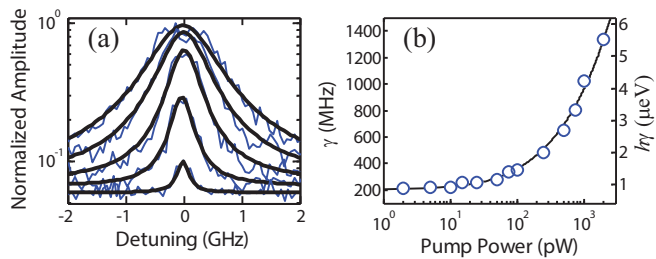}
\caption{Power broadening of the homogeneous linewidth for a sample temperature of 4.2 K and a probe power of 2 pW.  The homogeneous linewidth is obtained from a Lorentzian fit to the differential transmission lineshape, shown in (a) on a logarithmic vertical scale for increasing pump power from 2 pW to 500 pW.  (b) The linewidth data (open symbols) are fit with Eqn. \ref{eqn_powerbroadening} (solid line), yielding a zero-power linewidth of $\gamma_0 = 225 \pm 7$ MHz ($h\gamma_0 = 0.93 \pm 3$ $\mu$eV) and a saturation power $P_0 = 50 \pm 10$ pW.}
\label{fig4}
\end{figure}

To quantitatively establish the effects of the electron hyperfine interaction on the $X^+$ dephasing rate, we first perform saturation spectroscopy experiments in which $\gamma$ is measured as the pump power is increased from 2 pW to 2 nW in the absence of an external magnetic field.  Differential transmission lineshapes for increasing pump power are shown in Fig. \ref{fig4}(a).  For all pump powers, the lineshapes are fit well by a single Lorentzian.  We note that the Lorentzian fit is numerically superior to a Voigtian profile expected in the presence of inhomogeneous contributions to $\gamma$.  The half-width at half-maximum (HWHM) of the Lorentzian fit provides the dephasing rate, $\gamma$, shown in Fig. \ref{fig4}(b) by the open symbols.  With increasing pump power, $\gamma$ increases monotonically and is well described by a power broadening model of a two-level system \cite{Siegman1986,Allen1987}
\begin{equation}
\gamma = \gamma_0 \cdot \left(1 + P/P_0\right)^{1/2},
\label{eqn_powerbroadening}
\end{equation}
where $\gamma_0$ is the zero-power dephasing rate, $P$ is the pump power, and $P_0$ is the saturation pump power.  The fit of Eqn. \ref{eqn_powerbroadening} to the data is shown in Fig. \ref{fig4}(b) as the solid line, which yields a zero-power linewidth of $\gamma_0 = 225 \pm 7$ MHz ($h\gamma_0 = 0.93 \pm 3$ $\mu$eV) and a saturation power of $P_0 = 50 \pm 10$ pW.  Estimated uncertainties for $\gamma$ are on the order of $\pm 10$ MHz, which are too small to appear in the figure.

An advantage of the differential transmission experiment is the ability to distinguish between different dephasing mechanisms in the system, since different physical processes appear with unique lineshapes.  For example, carrier-phonon interactions at elevated sample temperatures will appear as a double Lorentzian lineshape with a narrow peak arising from the zero-phonon line superimposed onto a broad phonon background \cite{Palinginis2003}.  Phonon interactions are suppressed at 4.2 K, which is reflected in the single Lorentzian lineshapes in Fig. \ref{fig4}(a).  The sub-$\mu$eV linewidth at lower power also demonstrates that the differential transmission technique is insensitive to spectral diffusion effects or that they are absent from the sample.  The single Lorentzian lineshape and sub-$\mu$eV linewidth are consistent with the results reported by Cesari \textit{et al.} in which a low-temperature dephasing rate of 0.91 $\mu$eV was measured for a $p$-doped InAs/GaAs sample using time-integrated four-wave mixing\cite{Cesari2010}.  In that work, it was suggested that $\gamma$ was limited only by the population relaxation rate.  We demonstrate in the next subsection that in addition to spontaneous recombination at a rate $\Gamma_0$, the electron hyperfine interaction, which can increase $\Gamma$ beyond $\Gamma_0$ and introduce a non-zero $\gamma^*$, provides a lower limit for $\gamma$ in the absence of external fields.

\subsection{Magnetic Field Effects on the Charged Exciton: Faraday Configuration}

\begin{figure}[htb]
\centering
\includegraphics[width=0.95\columnwidth]{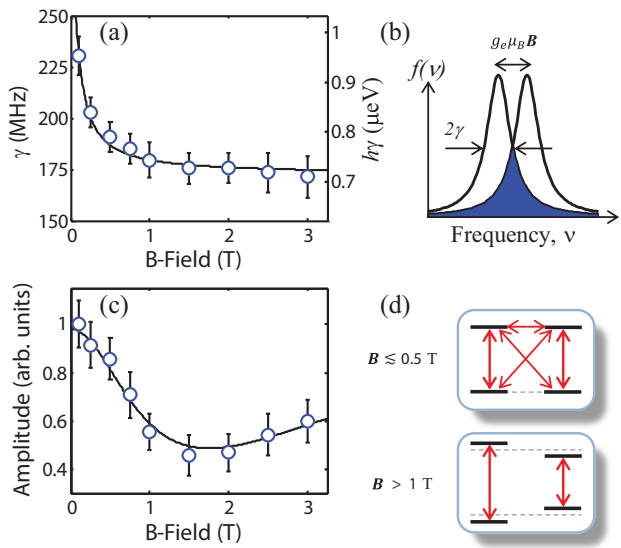}
\caption{(a) Dephasing rate, $\gamma$, as a function of applied magnetic field in the Faraday configuration.  Broadening beyond the radiative limit for applied field $\lesssim 1$ T is explained by a simple model shown in (b) describing the electron-nuclear coupling, which assumes that the strength of the hyperfine interaction is linearly related to the overlap of the charged exciton state distribution functions (given by the shaded area) separated in frequency by $g_e \mu_B \textbf{\textit{B}}$.  Each distribution is described by a Lorentzian function with HWHM linewidth $\gamma$.  The solid line in (a) is Eqn. \ref{eqn_width_HF} fit to the data with the fitting parameters $\Gamma^0_{\textrm{HF}} = 174$ MHz and $\gamma = 169$ MHz.  The large-field asymptote corresponds to the intrinsic radiatively-limited dephasing rate $\gamma = \Gamma_0/2 = 169$ MHz (0.70 $\mu$eV).  The normalized differential transmission amplitude is shown in (c) as the open symbols.  The solid line is Eqn. \ref{eqn_dT} fit to the data.  At \textbf{\textit{B}} $\approx 0$ T, bidirectional optical spin pumping inhibits shelving of the hole into either spin state, resulting in the maximum transmission amplitude.  For increasing \textbf{\textit{B}} up to $\approx 1.5$ T, the signal amplitude decreases as bidirectional spin pumping becomes inhibited by the broken degeneracy of the $X^+$ transitions.  At \textbf{\textit{B}} $\gtrsim 1.5$ T, as the optical spin pumping rate $\Gamma_{osp}$ decreases, spin pumping is suppressed, resulting in a slight recovery of the signal amplitude.  In (d), the dipole selection rules and dominant optical transitions in the QDs for weak and strong external field are shown.}
\label{fig5}
\end{figure}

The effects of the nuclear spin bath on the $X^+$ dephasing rate are investigated by performing differential transmission experiments with a non-zero external magnetic field in the Faraday configuration.  The pump and probe are both collinearly polarized along the $x$-direction with an average power of 10 pW and 2 pW, respectively.  The dependence of the dephasing rate on the external magnetic field is shown in Fig. \ref{fig5}(a) by the open symbols.  With increasing field from 0.1 T to 3 T, the dephasing rate decreases by $25 \%$, from $\gamma = 230$ MHz to $172$ MHz ($0.95$ $\mu$eV to $0.71$ $\mu$eV).

We interpret the magnetic field dependence as arising from the electron hyperfine interaction that enhances the dephasing rate for weak external fields and is suppressed for fields $\gtrsim 1$ T.  Since the hole hyperfine interaction is at least an order of magnitude weaker than for the electron, we consider here only electron-nuclear coupling.  At zero external magnetic field \textbf{\textit{B}}, the optically-excited electron spin in the $X^+$ state coherently precesses about the in-plane component of $\delta\textbf{\textit{B}}_N$, resulting in a non-zero $\gamma_e$ and $\Gamma_s$.  The dephasing rate associated with the hyperfine contributions relative to the $X^+$ spontaneous emission rate determines the dominant decay pathway.  Since the pump is resonant with both the $\sigma+$ and $\sigma-$ transitions, the differential probe transmission signal provides the ensemble-averaged dephasing rate for both transitions.

For non-zero \textbf{\textit{B}}, the degeneracy of the $X^+$ transitions is lifted as the energy of the $\sigma +$ ($\sigma -$) transition increases (decreases), which tunes QDs initially resonant with the laser for zero external field out of resonance.  For $\textbf{\textit{B}}\gg \gamma$, the laser is resonant with only a single transition in any particular QD -- the $\sigma +$ or $\sigma -$ transition for QDs initially redshifted or blueshifted, respectively, from the laser frequency.  In each QD, the electron spin precesses about the total magnetic field \textbf{\textit{B}}$_{{T}}$ = \textbf{\textit{B}} + $\delta\textbf{\textit{B}}_{{N}}$, which is aligned primarily along the $z$-direction for \textbf{\textit{B}} $\gg$ $\delta\textbf{\textit{B}}_{{N}}$.  A large external field stabilizes the electron spin, effectively screening it from $\delta\textbf{\textit{B}}_{{N}}$.  In this limit, hyperfine contributions to the $X^+$ dephasing rate are strongly suppressed so that $\gamma$ is limited only by $\Gamma_0$ and by the valence band mixing contribution, which is independent of the external magnetic field\cite{Dreiser2008,Kroner2008}.


To quantitatively support our interpretation of the field dependence of $\gamma$, we apply a model in which the strength of the hyperfine interaction, and thus the amplitude of electron spin precession about $\delta\textbf{\textit{B}}_N$ in the mean-field approximation, is determined by the probability to find the electron spin states at similar energy \cite{Kroner2008}.  For ground state electron spins in negatively charged QDs, Kroner \textit{et al.} demonstrated that the hyperfine interaction strength depends linearly on the energetic overlap of the electron spin state distribution functions, each given by a Lorentzian broadened by the finite spin state lifetime \cite{Kroner2008}.  In a positively charged QD, the electron forms a three-particle many-body state with the singlet hole pair, so that the Lorentzian distribution function is broadened not by the electron spin state lifetime, but by the $X^+$ dephasing rate, as shown in Fig. \ref{fig5}(b).  The distribution function is given by
\begin{equation}
f\left(\nu\right) \propto \frac{2}{\pi}\cdot\frac{\gamma}{4\left(\nu \pm g_e \mu_B \textbf{\textit{B}}/2\right)^2 + \left(\gamma\right)^2},
\label{eqn_electronbroadening}
\end{equation}
where the $X^+$ state splitting, $g_e\mu_B \textbf{\textit{B}}$, is determined by the electron Zeeman interaction since the singlet hole pair has zero net spin.  Electron-nuclear coupling, which requires conservation of both energy and angular momentum, can only occur for electrons with opposite spin and similar energy, since the nuclear spin state Zeeman splitting is orders of magnitude smaller than the splitting for electrons.  The probability of an electron spin flip is then given by the ratio of the overlap area of the distribution functions to their total area.  In the differential transmission experiments, the dephasing rates $\gamma_e$ and $\Gamma_s$ associated with the the $\left|\Uparrow\Downarrow\uparrow\right> \leftrightarrow \left|\Downarrow\Uparrow\downarrow\right>$ coherence and the ``forbidden" diagonal transitions, respectively, cannot be differentiated.  However, since both of these processes arise from the same hyperfine interaction mechanism, for simplicity we describe them using a single effective relaxation rate, $\Gamma_{\textrm{HF}}$, so that $\gamma = \left(\Gamma_0 + \Gamma_{\textrm{HF}}\right)/2$.  The hyperfine contribution to the $X^+$ dephasing rate, determined from the overlap of the distribution functions, is given by\cite{Kroner2008}
\begin{equation}
\Gamma_{\textrm{HF}} = \Gamma^0_{\textrm{HF}} \cdot \left(1 - \frac{2}{\pi}arctan\left(\frac{g_e \mu_B \textbf{\textit{B}}}{\gamma}\right)\right),
\label{eqn_width_HF}
\end{equation}
where $\Gamma^0_{\textrm{HF}}$ is the zero-field hyperfine-induced spin flip rate.  The result of the fit of Eqn. \ref{eqn_width_HF} to the data is shown in Fig. \ref{fig5}(a) by the solid line using a constant distribution linewidth of $\gamma = 169$ MHz.  Although the dephasing rate decreases by $\approx 25 \%$ at large external field, we obtain excellent agreement between the model and the experiment using constant broadening \cite{EN3} of the distribution functions for all applied \textbf{\textit{B}}.  From the fit, we obtain $\Gamma^0_{\textrm{HF}} = 174$ MHz, which corresponds to the zero-field electron-nuclear coupling rate that broadens $\gamma$ by $\Gamma^0_{HF}/2 = 87$ MHz beyond the radiative limit.  At large \textbf{\textit{B}}, the hyperfine interaction is suppressed and the fitted dephasing rate is limited only by the spontaneous recombination rate, $\gamma = \Gamma_0/2 = 169 \pm 7$ MHz ($0.70 \pm 0.03$ $\mu$eV).  The measured decrease of $\gamma$ by $\approx 60$ MHz (0.25 $\mu$eV) is consistent with the work of Kroner \textit{et al.}, in which the forbidden transition decay rate in a negatively charged QD, also mediated by the electron hyperfine interaction, decreased by $\approx 50$ MHz for $\textbf{\textit{B}} \approx 3$ T in the Faraday configuration\cite{Kroner2008}.

Strong hyperfine coupling of the electron spin states for weak external field strength also enables optical spin pumping (OSP), which is a key element for spin state initialization in single QD devices\cite{Gerardot2008,Xu2007}.  For $\sigma+$ or $\sigma-$ polarized optical excitation, the hole spin will be shelved into either the $\left|\Uparrow\right>$ or $\left|\Downarrow\right>$ states, respectively, as long as the spin pumping rate, $\Gamma_{osp}$, is faster than the hole spin state dephasing rate, $\gamma_h$.  We define the OSP fidelity as $\vartheta = \left(n_{\Uparrow} - n_{\Downarrow}\right)/\left(n_{\Uparrow} + n_{\Downarrow}\right)$, where $n_{\Uparrow}$ $\left(n_{\Downarrow}\right)$ is the population in the $\left|\Uparrow\right>$ $\left(\left|\Downarrow\right>\right)$ state.  $\vartheta$ can be related to $\Gamma_{osp}$ and $\gamma_h$ through a rate equation analysis\cite{Kroner2008} to give $\vartheta \approx \left(1 + 4\gamma_h/\Gamma_{osp}\right)^{-1}$.  Since the differential transmission amplitude is sensitive to the population of the hole spin ground states, this expression is used to estimate the signal amplitude as $\Delta T/T \approx \Delta T/T|_0\left(1 - \vartheta\right)$, where $\Delta T/T|_0$ is the maximum transmission signal when OSP is suppressed at large \textbf{\textit{B}}.  From Eqn. \ref{eqn_width_HF}, we can estimate the magnetic field dependence of the OSP rate as $\Gamma_{osp} \approx \textbf{\textit{B}}^{-2}$, so that $\vartheta \approx \left(1 + \gamma'\textbf{\textit{B}}^2\right)^{-1}$, where $\gamma'$ is a phenomenological fitting parameter that takes into account both $\gamma_h$ and $\Gamma_{osp}$ at zero field.  At large \textbf{\textit{B}}, $\vartheta \rightarrow 0$ and $\Delta T/T \rightarrow \Delta T/T|_0$.  For circularly polarized excitation of either the $\sigma+$ or the $\sigma-$ $X^+$ transition, as \textbf{\textit{B}} $\rightarrow 0$, $\Gamma_{osp}$ $\gg \gamma_h$ and $\Delta T/T \rightarrow 0$.

For $x$-polarized optical excitation used in our work, OSP is frustrated at small \textbf{\textit{B}} since the pump is quasi-resonant with both $X^+$ transitions.  In this case, the $\sigma+$ ($\sigma-$) component of the linearly polarized pump acts as both an excitation field for OSP to the $\left|\Uparrow\right>$ ($\left|\Downarrow\right>$) state and as a re-pump field to frustrate OSP to the $\left|\Downarrow\right>$ ($\left|\Uparrow\right>$) state arising from the $\sigma-$ ($\sigma+$) polarized component.  As a result, instead of $\Delta T/T \rightarrow 0$ at zero field, bidirectional OSP drives $\Delta T/T \rightarrow \Delta T/T|_0$. Two-color resonant Rayleigh scattering experiments \cite{Gerardot2008} have shown that the re-pumping efficiency decreases as the re-pump field is detuned from the $X^+$ transition by an energy $\Delta E$.  As a result, the bidirectional OSP contribution to $\Delta T/T$ decreases as $\approx \Delta E^{-2}$, where $\Delta E = \left(g_e + g_h\right) \mu_B \textbf{\textit{B}}$ in our experiments.  We can approximate the total normalized differential transmission signal as a sum of the OSP and re-pumping contributions:
\begin{equation}
\frac{\Delta T/T}{\Delta T/T|_0} \approx \left[\left(\frac{1}{1+\textbf{\textit{B}}^2}\right) + \left(1 - \frac{1}{1 + \gamma'\textbf{\textit{B}}^2}\right)\right],
\label{eqn_dT}
\end{equation}
where the first term arises from the re-pumping process and the second term from OSP.  We use $1/\left(1 + \textbf{\textit{B}}^2\right)$ for re-pumping instead of $1/\textbf{\textit{B}}^2$ to take into account the fact that at zero field, the normalized signal amplitude is unity for maximum re-pumping efficiency.  The fit of Eqn. \ref{eqn_dT} to the transmission amplitude data using $\gamma' = 0.025$ is given by the solid line in Fig. \ref{fig5}(c).

Despite the simplicity of the model, excellent agreement with the measurements indicates that the model captures the essential processes responsible for the amplitude dependence on \textbf{\textit{B}}.  At small \textbf{\textit{B}}, OSP is frustrated by the re-pumping field so that the two terms nearly cancel and the normalized transmission amplitude is $\approx 1$.  With increasing \textbf{\textit{B}}, the Zeeman splitting of the electron and hole spin states breaks the degeneracy of the optically active $X^+$ transitions.  A larger $\Delta E$ reduces the rate at which the $x$-polarized excitation field can re-pump the shelved hole spin so that the normalized transmission amplitude decreases for \textbf{\textit{B}} up to $\sim 1.5$ T.  For \textbf{\textit{B}} $\gtrsim 1.5$ T, the small OSP rate is reflected in the partial recovery of the transmission amplitude.  To further investigate the magnetic field dependence of the dephasing rate, in the next subsection we use a field in the Voigt configuration to mix the spin states, which opens additional dephasing channels for $X^+$.

\subsection{Magnetic Field Effects on the Charged Exciton: Voigt Configuration}

\begin{figure}[htb]
\centering
\includegraphics[width=0.95\columnwidth]{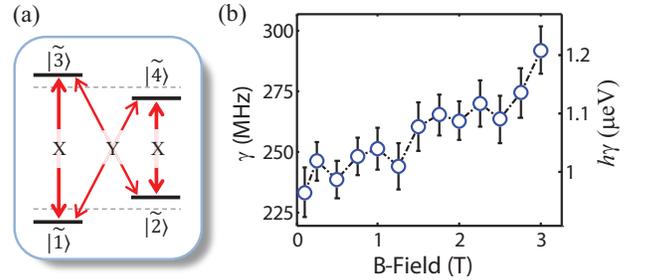}
\caption{(a) The dipole selection rules and dominant optical transitions in the QD under strong applied field in the Voigt configuration, where $X$ and $Y$ are the dipole moments aligned along and perpendicular to the Voigt field, respectively. The dephasing rate, $\gamma$, as a function of magnetic field is shown in (b).}
\label{fig6}
\end{figure}

For a magnetic field in the Voigt configuration ($\theta = 0$ in Fig. \ref{fig1}(a)), the spins are quantized along the $x$-direction parallel to the applied field.  In the $z$-basis, the Voigt field mixes the spin states and the ``forbidden" transitions become optically active.  For \textbf{\textit{B}} $\gtrsim 1$ T where the hyperfine interaction is weakest, the hole and charged exciton states form a four-level system with linearly polarized optical selection rules, shown in Fig. \ref{fig6}(a).  The ground and excited states can be written as linear combinations of the hole spin states $\tilde{\left|1\right>} = \left(\left|\Uparrow\right> + \left|\Downarrow\right>\right)/\sqrt{2}$ and $\tilde{\left|2\right>} = \left(\left|\Uparrow\right> - \left|\Downarrow\right>\right)/\sqrt{2}$ and the charged exciton states $\tilde{\left|3\right>} = \left(\left|\Uparrow\Downarrow\uparrow\right> + \left|\Downarrow\Uparrow\downarrow\right>\right)/\sqrt{2}$ and $\tilde{\left|4\right>} = \left(\left|\Uparrow\Downarrow\uparrow\right> - \left|\Downarrow\Uparrow\downarrow\right>\right)/\sqrt{2}$, respectively.  Transitions between $\tilde{\left|1\right>} \leftrightarrow \tilde{\left|3\right>}$ and $\tilde{\left|2\right>} \leftrightarrow \tilde{\left|4\right>}$ are linearly polarized along the $x$-direction.  Transitions between $\tilde{\left|1\right>} \leftrightarrow \tilde{\left|4\right>}$ and $\tilde{\left|2\right>} \leftrightarrow \tilde{\left|3\right>}$ are polarized along the $y$-direction and are not accessible using the waveguide geometry in the absence of heavy-hole--light-hole mixing.

For a weak magnetic field, the hyperfine interaction dominates over state mixing and the optical transitions are circularly polarized.  In this regime, $\gamma = 235 \pm 7$ MHz ($0.97 \pm 3$ $\mu$eV), as shown in Fig. \ref{fig6}(b).  As expected, $\gamma$ is similar for the Voigt and Faraday configurations at small $\textbf{\textit{B}}$.  Similar to the Faraday configuration experiments, with increasing \textbf{\textit{B}}, the degeneracy of the $X^+$ transitions is lifted and QDs initially resonant with the pump laser at $\textbf{\textit{B}} = 0$ T are shifted out of resonance.  At large \textbf{\textit{B}}, the $\tilde{\left|1\right>} \leftrightarrow \tilde{\left|3\right>}$ $\left(\tilde{\left|2\right>} \leftrightarrow \tilde{\left|4\right>}\right)$ transition for QDs initially redshifted (blueshifted) from the laser frequency is tuned into resonance by the magnetic field, and the pump and probe are resonant with only one of these transitions in a given QD.  In contrast to the Faraday experiments, however, the $X^+$ dephasing rate \textit{increases} to $\gamma = 293 \pm 9$ MHz ($1.21 \pm 0.04$ $\mu$eV) at $\textbf{\textit{B}} = 3$ T.

The observed behavior of $\gamma$ with increasing \textbf{\textit{B}} is a result of the field-induced state mixing, which enhances the dipole moment of the diagonal transitions.  For $\textbf{\textit{B}} \gg \delta\textbf{\textit{B}}_N$, hyperfine interaction contributions to the dephasing rate are suppressed and the eigenstates of the system are the basis states of the Voigt field Hamiltonian.  In the radiative limit, the dephasing rate of a charged exciton in either state $\tilde{\left|3\right>}$ or $\tilde{\left|4\right>}$ is given by $\gamma = \left(\Gamma^X_0 + \Gamma^Y_0\right)/2$, where $\Gamma^X_0$ and $\Gamma^Y_0$ are the spontaneous recombination rates for the $X$ and $Y$ transitions, respectively.  If we assume $\Gamma^X_0 = \Gamma^Y_0 = \Gamma_0$, then we expect $\gamma = \Gamma_0 = 339$ MHz at large \textbf{\textit{B}}.  This value is comparable to the measured $\gamma = 293 \pm 9$ MHz at $\textbf{\textit{B}} = 3$ T.  The discrepancy between the measured and expected values is likely due to unequal dipole moments of the $X$ and $Y$ polarized transitions.  For example, if we assume the $X$ polarized transition spontaneous recombination rate $\Gamma^X_0 = \Gamma_0 = 339$ MHz, then from the measured $\gamma = 293$ MHz, we get $\Gamma^Y_0 = 2\gamma - \Gamma^X_0 = 247$ MHz, which corresponds to a $Y/X$ transition dipole moment ratio\cite{EN4} $d_Y/d_X = \sqrt{\Gamma^X_0/\Gamma^Y_0} \approx 0.85$.

The large dipole moment of the diagonal transitions enables OSP with an efficiency increasing with \textbf{\textit{B}}, in contrast to the behavior in the Faraday configuration.  For $x$-polarized optical excitation resonant with, e.g. the $\tilde{\left|1\right>} \leftrightarrow \tilde{\left|3\right>}$ transition, the hole will be pumped into state $\tilde{\left|2\right>}$ via recombination along $\tilde{\left|3\right>} \leftrightarrow \tilde{\left|2\right>}$ with the emission of a $y$-polarized photon.  As long as the hole spin flip rate $\gamma_h$ between states $\tilde{\left|1\right>}$ and $\tilde{\left|2\right>}$ is negligible compared to $\Gamma_{osp}$, a significant hole spin population will build up in state $\tilde{\left|2\right>}$.  The OSP fidelity is partially limited by re-pumping along $\tilde{\left|2\right>} \leftrightarrow \tilde{\left|4\right>}$ by the excitation laser, which is detuned from the transition by an energy $\Delta E = \left(g^X_e + g^X_h\right)\mu_B\textbf{\textit{B}}$, where $g^X_e$ and $g^X_h$ are the in-plane electron and hole g-factors, respectively.  The OSP fidelity for this configuration has been shown to depend on $\Delta E$ according to $\vartheta \approx 1 - \left(\Gamma^2_0 + 3\Omega^2\right)/\Delta E^2$, where $\Omega$ is the Rabi frequency of the optical excitation field \cite{Emary2007}.  With increasing \textbf{\textit{B}}, $\Delta E$ increases and $\vartheta \rightarrow 1$.  As a result, the differential transmission signal amplitude $\Delta T/T$ should decrease with increasing \textbf{\textit{B}}.  This is indeed what we observe for $\textbf{\textit{B}} \gtrsim 1.5$ T (data not shown).  For $\textbf{\textit{B}} \lesssim 1.5$ T, $\Delta T/T$ exhibits a non-trivial dependence on the magnetic field which suggests that the optical selection rules are not strictly linear and that the hyperfine interaction plays a role.  Additionally, heavy-hole--light-hole mixing caused by the QD in-plane anisotropy can rotate the polarization axis of the QD away from the applied field so that the $X$ and $Y$ transitions are neither parallel nor perpendicular to the external field\cite{Koudinov2004,Xu2007}.  In this case, the $x$-polarized pump field can excite all four transitions, which would enhance the re-pumping efficiency along the diagonal transitions that have a smaller $\Delta E$.

\section{Conclusions}

Contributions to the positively charged exciton dephasing rate from the electron hyperfine interaction in self-assembled InAs QDs have been investigated using high resolution two-color differential transmission spectroscopy.  We demonstrate that in the absence of carrier- and phonon-mediated pure dephasing effects, the hyperfine interaction is the dominant dephasing mechanism and introduces a lower bound on the charged exciton dephasing rate for the case of zero applied magnetic field.  The strong electron hyperfine interaction can flip the electron spin, coupling the charged exciton spin states and enabling the diagonal ``forbidden" transitions.  These additional decay pathways can be suppressed with the application of a moderate Faraday magnetic field to give a dephasing rate of $\gamma = 169$ MHz (0.70 $\mu$eV), which is limited only by spontaneous recombination along the dipole-allowed transitions.  The opposite is true for an applied Voigt magnetic field: the in-plane field mixes the spin eigenstates, enabling the ``forbidden" transitions and consequently enhancing the radiatively limited dephasing rate by $\approx 75 \%$.

The magnetic field dependent measurements of the dephasing rate presented in this work might help guide efforts for quantum communication and computation using semiconductor QDs.  For example, Yao, Liu, and Sham proposed a deterministic method for entangling two spatially separated QD qubits using an intermediate spin-photon entangled state \cite{Yao2005}.  Initialization of the qubit into a pre-entangled state was recently demonstrated using a positively charged QD, with a fidelity below unity partially due to radiative decay and pure dephasing \cite{Webster2014}.  The results presented here demonstrate that a higher fidelity might be achievable by tailoring external fields to minimize dephasing from electron-nuclear interactions.  Additional insight into excitonic hyperfine interactions could be obtained by measuring the population relaxation and pure dephasing rates of individual charged exciton transitions.  This could be achieved by performing additional experiments such as optical coherent multi-dimensional spectroscopy, which is specifically tailored to isolate quantum pathways and can readily separate pure dephasing from incoherent relaxation mechanisms \cite{Cundiff2012}.  This technique could be applied to an isolated QD embedded in the intrinsic region of a $p-i-n$ diode for photocurrent detection\cite{Betz2010,Nardin2013}, which would provide well-resolved and spectrally isolated signals associated with each dipole-allowed transition of the four-level hole--charged-exciton system.

\section*{Acknowledgments}

We thank Todd Harvey of the National Institute of Standards and Technology for growth of the epitaxial sample.

\bibliography{RefList1}{}

\begin{thebibliography}{45}
\bibitem [{EN1()}]{EN1}%
  \BibitemOpen
  \href@noop {} {\bibinfo  {journal} {Each Au/insulator/semiconductor interface
  of the waveguide forms a MIS-capacitor, so that the device acts as a
  dual-gate capacitor. One-dimensional Sch$\ddot{\textrm{o}}$dinger-Poisson
  equation calculations indicate that for the properties of the device under
  test, the dual-gate capacitance for one electrode at V = 0 is nearly
  identical to the single-gate capacitance (see Refs. [26, 27]). We therefore
  use a single-gate model to determine the doping density}\ }\BibitemShut
  {NoStop}%
\bibitem [{EN2()}]{EN2}%
  \BibitemOpen
  \href@noop {} {\bibinfo  {journal} {We use the static dielectric constant of
  Al$_{0.7}$Ga$_{0.3}$As since this material constitutes the majority of the
  waveguide}\ }\BibitemShut {NoStop}%
\bibitem [{EN3()}]{EN3}%
  \BibitemOpen
  \href@noop {} {\bibinfo  {journal} {We find no qualitative differences
  between the fits when using a constant $h\gamma$ between the measured range
  of dephasing rates $0.7$ $\mu$eV $\leq h\gamma \leq1$ $\mu$eV}\ }\BibitemShut
  {NoStop}%
\bibitem [{EN4()}]{EN4}%
  \BibitemOpen
\bibfield  {journal} {  }\href@noop {} {\bibinfo  {journal} {One possible
  explanation for $d_Y/d_X < 1$ could be a slight misalignment of the Voigt
  field with respect to the sample, which would reduce the state mixing. We
  estimate that $\theta = 0^{\circ} \pm 2.5^{\circ}$ in the Voigt
  configuration}\ }\BibitemShut {NoStop}%
\end{thebibliography}

\end{document}